\pdfoutput=1 
\documentclass[twocolumn]{aastex62}
\usepackage{amsmath,amstext}
\usepackage[T1]{fontenc}
\usepackage[figure,figure*]{hypcap}

\def\bfref{}

\begin{document}

\title{PALFA Discovery of a Highly Relativistic Double Neutron Star Binary}

\author{K.~Stovall}
\affiliation{National Radio Astronomy Observatory, 1003 Lopezville Road, Socorro, NM, 87801, USA}

\author{P.~C.~C.~Freire}
\affiliation{Max-Planck-Institut f\"{u}r Radioastronomie, Auf dem H\"{u}gel 69, 53131 Bonn, Germany}

\author{S.~Chatterjee}
\affiliation{Cornell Center for Astrophysics and Planetary Science and Department of Astronomy, Cornell University, Ithaca, NY 14853, USA}

\author{P.~B.~Demorest}
\affiliation{National Radio Astronomy Observatory, 1003 Lopezville Road, Socorro, NM, 87801, USA}

\author{D.~R.~Lorimer}
\affiliation{Dept. of Physics and Astronomy, West Virginia Univ., Morgantown, WV 26506, USA}
\affiliation{Center for Gravitational Waves and Cosmology, Chestnut Ridge Research Building, Morgantown, WV 26505}

\author{M.~A.~McLaughlin}
\affiliation{Dept. of Physics and Astronomy, West Virginia Univ., Morgantown, WV 26506, USA}
\affiliation{Center for Gravitational Waves and Cosmology, Chestnut Ridge Research Building, Morgantown, WV 26505}

\author{N.~Pol}
\affiliation{Dept. of Physics and Astronomy, West Virginia Univ., Morgantown, WV 26506, USA}
\affiliation{Center for Gravitational Waves and Cosmology, Chestnut Ridge Research Building, Morgantown, WV 26505}

\author{J.~van~Leeuwen}
\affiliation{ASTRON, the Netherlands Institute for Radio Astronomy, Postbus 2, 7990 AA, Dwingeloo, The Netherlands}
\affiliation{Anton Pannekoek Institute for Astronomy, Univ. of Amsterdam, Science Park 904, 1098, XH Amsterdam, The Netherlands}

\author{R.~S.~Wharton}
\affiliation{Max-Planck-Institut f\"{u}r Radioastronomie, Auf dem H\"{u}gel 69, 53131 Bonn, Germany}
\affiliation{Cornell Center for Astrophysics and Planetary Science and Department of Astronomy, Cornell University, Ithaca, NY 14853, USA}

\author{B.~Allen}
\affiliation{Max-Planck-Institut f\"{u}r  Gravitationsphysik,  D-30167  Hannover, Germany}
\affiliation{Leibniz Universit\"{a}t Hannover, D-30167 Hannover, Germany}
\affiliation{Physics Dept., Univ. of Wisconsin - Milwaukee, Milwaukee WI 53211, USA}

\author{M.~Boyce}
\affiliation{Dept. of Physics \& McGill Space Institute, McGill Univ., Montreal, QC H3A 2T8, Canada}

\author{A.~Brazier}
\affiliation{Cornell Center for Astrophysics and Planetary Science and Department of Astronomy, Cornell University, Ithaca, NY 14853, USA}
\affiliation{Cornell Center for Advanced Computing, Cornell University, Ithaca, NY 14850, USA}

\author{K.~Caballero}
\affiliation{Center for Advanced Radio Astronomy, Univ. of Texas Rio Grande Valley, Brownsville, TX 78520, USA}

\author{F.~Camilo}
\affiliation{SKA South Africa, Pinelands, 7405, South Africa}

\author{R.~Camuccio}
\affiliation{Center for Advanced Radio Astronomy, Univ. of Texas Rio Grande Valley, Brownsville, TX 78520, USA}

\author{J.~M.~Cordes}
\affiliation{Cornell Center for Astrophysics and Planetary Science and Department of Astronomy, Cornell University, Ithaca, NY 14853, USA}

\author{F.~Crawford}
\affiliation{Dept. of Physics and Astronomy, Franklin and Marshall College, Lancaster, PA 17604-3003, USA}

\author{J.~S.~Deneva}
\affiliation{George Mason University, resident at the Naval Research Laboratory, 4555 Overlook Ave. SW, Washington, DC 20375, USA}

\author{R.~D.~Ferdman}
\affiliation{Faculty of Science, University of East Anglia, Norwich Research Park, Norwich NR4 7TJ, United Kingdom}

\author{J.~W.~T.~Hessels}
\affiliation{ASTRON, the Netherlands Institute for Radio Astronomy, Postbus 2, 7990 AA, Dwingeloo, The Netherlands}
\affiliation{Anton Pannekoek Institute for Astronomy, Univ. of Amsterdam, Science Park 904, 1098, XH Amsterdam, The Netherlands}

\author{F.~Jenet}
\affiliation{Center for Advanced Radio Astronomy, Univ. of Texas Rio Grande Valley, Brownsville, TX 78520, USA}

\author{V.~M.~Kaspi}
\affiliation{Dept. of Physics \& McGill Space Institute, McGill Univ., Montreal, QC H3A 2T8, Canada}

\author{B.~Knispel}
\affiliation{Max-Planck-Institut f\"{u}r  Gravitationsphysik,  D-30167  Hannover, Germany}
\affiliation{Leibniz Universit\"{a}t Hannover, D-30167 Hannover, Germany}

\author{P.~Lazarus}
\affiliation{Max-Planck-Institut f\"{u}r Radioastronomie, Auf dem H\"{u}gel 69, 53131 Bonn, Germany}

\author{R.~Lynch}
\affiliation{NRAO, Charlottesville, VA 22903, USA}

\author{E.~Parent}
\affiliation{Dept. of Physics \& McGill Space Institute, McGill Univ., Montreal, QC H3A 2T8, Canada}

\author{C.~Patel}
\affiliation{Dept. of Physics \& McGill Space Institute, McGill Univ., Montreal, QC H3A 2T8, Canada}

\author{Z. Pleunis}
\affiliation{Dept. of Physics \& McGill Space Institute, McGill Univ., Montreal, QC H3A 2T8, Canada}

\author{S.~M.~Ransom}
\affiliation{NRAO, Charlottesville, VA 22903, USA}

\author{P.~Scholz}
\affiliation{National Research Council of Canada, Herzberg Astronomy and Astrophysics, Dominion Radio Astrophysical Observatory, P.O. Box 248, Penticton, BC V2A 6J9, Canada}

\author{A.~Seymour}
\affiliation{Arecibo Observatory, HC3 Box 53995, Arecibo, PR 00612}

\author{X.~Siemens}
\affiliation{Physics Dept., Univ. of Wisconsin - Milwaukee, Milwaukee WI 53211, USA}

\author{I.~H.~Stairs}
\affiliation{Dept. of Physics and Astronomy, Univ. of British Columbia, Vancouver, BC V6T 1Z1, Canada}

\author{J.~Swiggum}
\affiliation{Physics Dept., Univ. of Wisconsin - Milwaukee, Milwaukee WI 53211, USA}

\author{W.~W.~Zhu}
\affiliation{National Astronomical Observatories, Chinese Academy of Sciences, A20 Datun Rd, Chaoyang District, Beijing 100012, P. R. China}
\affiliation{Max-Planck-Institut f\"{u}r Radioastronomie, Auf dem H\"{u}gel 69, 53131 Bonn, Germany}

\correspondingauthor{K.~Stovall}
\email{kstovall@nrao.edu}

\begin{abstract}
We report the discovery and initial follow-up of a double neutron star
(DNS) system, PSR J1946$+$2052, with the Arecibo L-Band Feed Array pulsar (PALFA)
survey. PSR J1946$+$2052 is a 17-ms pulsar in a 1.88-hour, eccentric
($e \, =\, 0.06$) orbit {\bfref with a $\gtrsim 1.2 \, M_\odot$ companion.}
We have used the Jansky Very Large Array to localize PSR J1946$+$2052 to a precision
of $0\farcs09$ using a new phase binning mode. We have searched multiwavelength catalogs
for coincident sources but did not find any counterparts. The improved position enabled a
measurement of the spin period derivative of the pulsar
($\dot{P} \, = \, 9\,\pm \, 2 \,\times 10^{-19}$); the small inferred magnetic field strength at the surface
($B_S \, = \, 4 \, \times \, 10^9 \, \rm G$) indicates that this
pulsar has been recycled. This and the orbital eccentricity lead to the conclusion that
PSR J1946$+$2052 is in a DNS {\bfref system}. Among all known radio pulsars in DNS systems, PSR J1946$+$2052 has
the shortest orbital period and the shortest estimated merger timescale, 46 Myr; at that time
it will display the largest spin effects on gravitational wave waveforms of any such
system discovered to date.  We have measured the advance of periastron passage for this system,
$\dot{\omega} \, = \, 25.6 \, \pm \, 0.3\, \deg \rm yr^{-1}$, implying a total system mass
of only 2.50 $\pm$ 0.04 $M_\odot$, {\bfref so it is among the lowest mass DNS systems. This total
mass measurement combined with the minimum companion mass constrains the pulsar mass to $\lesssim 1.3 \, M_\odot$.}
\end{abstract}


\section{Introduction}\label{sec:intro}
Since the discovery of PSR~B1913+16 \citep{1975ApJ...195L..51H},
double neutron star (DNS) systems have
allowed a wide range of investigations into many aspects of astrophysics and fundamental physics.
Paramount among these have been tests of general relativity (GR) and alternative theories of gravity.
The exquisite match between the observed rate of orbital decay of PSR~B1913$+$16 and that predicted by GR due to the emission
of gravitational waves (GWs) \citep{1991ApJ...366..501D,2016ApJ...829...55W} 
showed that GR gives a self-consistent description of relativistic effects. Moreover, it established
experimentally that GWs are not a mere coordinate effect: they carry energy across space, 
and have a real effect on the orbital dynamics of massive objects. This  indirect
detection of GWs preceded the first direct detections~\citep{2016PhRvL.116f1102A} by decades. The
continued orbital decay in PSR~B1913+16 inevitably leads to the neutron stars merging
and  GWs from such a merger have recently been detected \citep{2017PhRvL.119p1101A}.

Twelve more DNS systems have since been discovered, with 
another three DNS candidate systems unconfirmed (for a review, see 
\citealt{2017ApJ...846..170T}). Several of these---PSRs J0737$-$3039A/B \citep{2003Natur.426..531B}, B1534+12 \citep{2014ApJ...787...82F},
J1756$-$2251 \citep{2014MNRAS.443.2183F}, J1757$-$1854 \citep{2017arXiv171107697C},
J1906+0746
\citep[another PALFA discovery;][]{2006ApJ...640..428L,2015ApJ...798..118V}, and B2127+11C \citep{2006ApJ...644L.113J}---have also been used to test the
predictions of theories of gravity.

Of these, J0737$-$3039A/B, the ``double pulsar'' has been the most outstanding test system.
The discovery of the recycled pulsar in the system, PSR~J0737$-$3039A, was in itself sufficient to
significantly increase the estimated Galactic DNS merger rate \citep{2003Natur.426..531B,2004ApJ...601L.179K}.
The discovery that the second NS in the system is also a pulsar, PSR~J0737$-$3039B
\citep{2004Sci...303.1153L}, allowed for a total of four independent and stringent
tests of GR from timing observations alone \citep{2006Sci...314...97K}.

Given the extraordinary scientific results that have emerged from the study of DNS systems,
their discovery has been an important motivation for many ongoing pulsar surveys.
In this paper, we focus on a discovery from the PALFA survey
\citep{2006ApJ...637..446C,2015ApJ...812...81L}, currently being carried out with the Arecibo Observatory.
PALFA has thus far resulted in the discovery of 180 pulsars, including
22 millisecond pulsars \citep[e.g.][]{2012ApJ...757...89D,2013ApJ...773...91A,2015ApJ...800..123S,2015ApJ...806..140K,2016ApJ...833..192S}, 
two DNS systems, PSRs J1906+0746 and J1913+1102 \citep{2006ApJ...640..428L,2016ApJ...831..150L},
and a repeating fast radio burst (FRB)~\citep{2016Natur.531..202S}. In this letter, we present the discovery 
of PSR J1946$+$2052, a 17-ms pulsar in
a 1.88-hour, eccentric ($e \, = \, 0.064$) orbit with a $\sim$1.2 $M_\odot$ companion. This
is the third DNS system found in PALFA {\bfref and is the DNS system with the shortest orbital period}.

\section{Observations and Analysis}\label{sec:obs}
The PALFA survey uses the Arecibo L-Band Feed Array (ALFA) receiver's 7 beams to
search the Galactic plane ($b < |5^\circ|$) visible by
the Arecibo Observatory for pulsars and FRBs. The
survey consists of two portions, on the inner 
($32^\circ < l < 77^\circ$) and outer 
($168^\circ < l < 214^\circ$) Galaxy.
These are identical in setup except for the pointing integration time of 260\,s for the inner versus 180\,s for the outer Galaxy. The survey uses the
Mock spectrometers centered at 1375.489\,MHz over 322.398\,MHz of bandwidth
divided into 960 frequency channels, sampled every 65\,$\mathrm{\mu s}$~\citep{2015ApJ...812...81L}. 

\subsection{Discovery and early follow-up}\label{subsec:disc}
The PALFA survey identifies candidate discoveries using 3 separate pipelines:
1) a reduced time-resolution pipeline using the {\tt PRESTO} software suite \citep[][]{2001PhDT.......123R} without
acceleration searching known as the `Quicklook' pipeline~\citep{2013PhDT.......127S}, 2) a
full-resolution {\tt PRESTO} pipeline with enhanced radio frequency interference (RFI) mitigation techniques and
searches for acceleration up to 1650 $\mathrm{m\;s^{-2}}$ for a 10-ms pulsar~(see~\citealt{2015ApJ...812...81L}),
and 3) an Einstein@Home pipeline which searches for tight binaries using a template-matching
search~\citep{2013ApJ...773...91A}.

PSR J1946$+$2052 was discovered in pipeline 1), so we will briefly describe it here. The Quicklook pipeline is run
on-site at the Arecibo Observatory to rapidly identify strong pulsar signals. Data from
the Mock spectrometers are converted from 8-bit to 4-bit and the two sub-bands for each beam are combined into a single {\tt PSRFITS}
file as described in Section~3.2 of~\cite{2015ApJ...812...81L}. Each combined file is reduced in time-resolution by a factor of two 
and is examined by {\tt rfifind} in order to create a RFI mask. The data are then de-dispersed at a series of trial
DMs ranging from 0 to 1,550.5 $\mathrm{pc\;cm^{-3}}$ and searched for periodic signals using {\tt accelsearch} summing up
to 16 harmonics, but without searching for acceleration. The candidates are sorted by significance and the top 20 candidates
are folded into diagnostic plots. The plots are examined using the PEACE algorithm~\citep{2013MNRAS.433..688L}
to identify the most promising candidates.

\begin{figure}
\begin{center}
\includegraphics[width=\columnwidth]{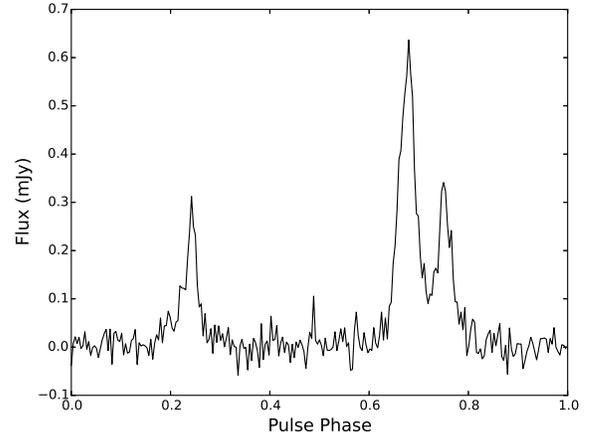}
\caption{Average profile for PSR J1946+2052 at 1.43\,GHz from a 2-hour observation {\bfref from the Arecibo Observatory} using the PUPPI backend in coherent search mode.}
\label{fig:profile}
\end{center}
\end{figure}

In PALFA observations recorded on 2017 July 19, inspection of the Quicklook pipeline diagnostic plots resulted in the discovery 
of a 17-ms pulsar with a significant apparent period 
change within the 260-s observation at a DM of about 94 $\mathrm{pc\;cm^{-3}}$. Subsequent Arecibo observations in 2017 August
and September were
performed using the L-wide receiver and the PUPPI backend configured in coherently dedispersed search mode. The PUPPI observations were
recorded at a center frequency of 1381 MHz with a bandwidth of 800 MHz across 512 frequency channels that were coherently de-dispersed
at the pulsar's best known DM at the time of observation. Samples were recorded every 10.24 $\mathrm{\mu s}$. We used some
of these initial observations to determine the pulsar's orbit using {\tt PRESTO}'s {\tt fitorb.py} and created a
preliminary ephemeris. Observations were folded using this ephemeris into 10-s subintegrations, cleaned of RFI,
and reduced to two frequency channels and 60-s
subintegrations. The pulse profile for PSR J1946$+$2052 from a 2-hr observation is shown in Figure~\ref{fig:profile}.
This observation has been polarization- and flux- calibrated 
by scaling the pulsar observation using an observation of a noise diode injected signal and an unpolarized quasar (J1445+0958).
No polarization has been detected in PSR J1946$+$2052, so we only show total intensity.
We generated a template for the characteristic pulse shape of PSR J1946+2052 by summing 40 min of observations
together and smoothing the summed profile. Times-of-arrival (TOAs) were then generated by 
cross-correlating each frequency channel for every 60-s subintegration in the Fourier domain~\citep{1992PTRSL.341..117T}
with our template using the {\tt PSRCHIVE} tool {\tt pat}~\citep{2004PASA...21..302H}. The resulting TOAs were compared to a model for PSR J1946$+$2052 using
the {\tt tempo} pulsar timing software. 

On 2017 September 28, we used the Robert C. Byrd Green Bank Telescope to
observe PSR J1946$+$2052 at 820 MHz using GUPPI~\citep{2008SPIE.7019E..1DD} with a bandwidth of 200 MHz across 128 frequency channels in coherent
search mode with a sample time of 10.24 $\mathrm{\mu s}$. We used the same procedure as above to obtain TOAs.
To properly account for the time offset between the GBT-GUPPI observations and
Arecibo-PUPPI observations, we used offsets measured by the NANOGrav collaboration from observations of PSR
J1713$+$0747~\citep{2015ApJ...813...65T}.

Since we have been recording follow-up data in coherent search mode with GUPPI and PUPPI, this has enabled for us to search for a potential companion pulsar.
Thus far we have not detected the companion as a pulsar.

\subsection{Localization}

\begin{figure*}
\includegraphics[width=\textwidth]{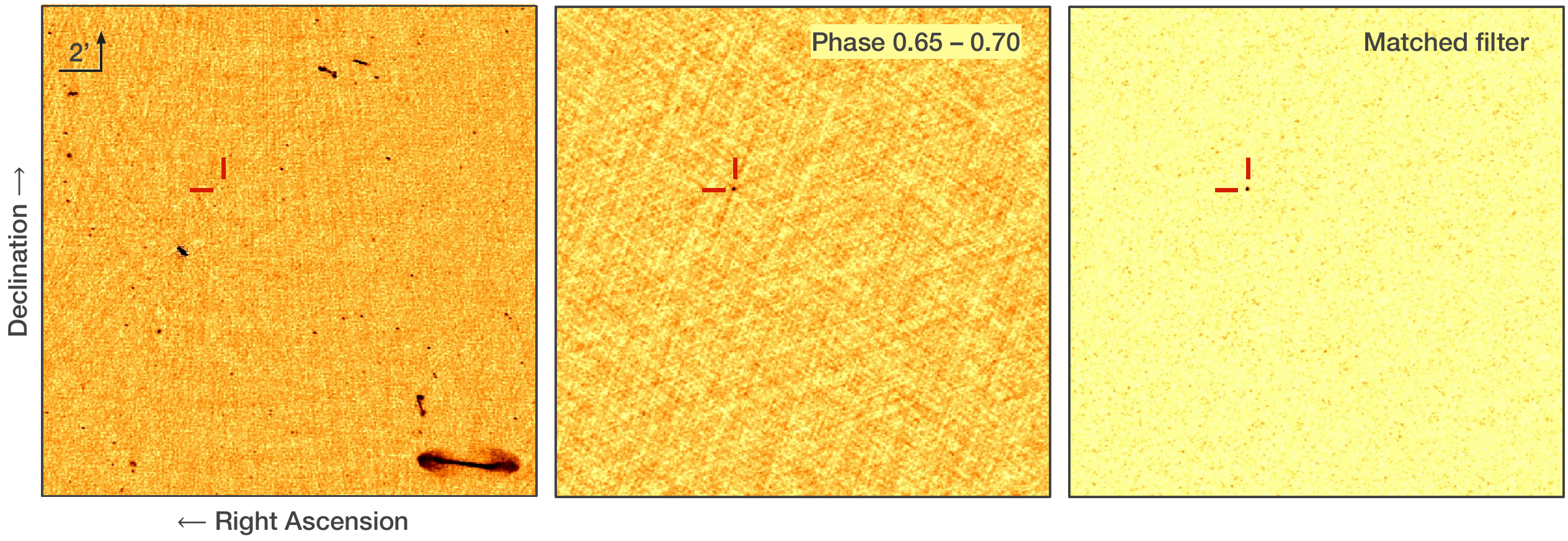}
\caption{Left: Bin-averaged image from 1 to 2 GHz of the field with PSR J1946$+$2052.
Center: Zoomed-in version of left plot showing only one of the phase bins; here the pulsar becomes obvious.
Right: Image showing the maximum of the convolution with the template profile for each pixel.}
\label{fig:VLA}
\end{figure*}

ALFA has a beam size (FWHM) of 3\farcm35 \citep{2006ApJ...637..446C}; this represents
the approximate uncertainty in the sky location of the pulsar at the time of discovery. To better localize the pulsar, we observed its approximate position
using the Karl G. Jansky Very Large Array (VLA) during the move from C to B configuration, however most
of the antennas were already in their locations for B configuration. We observed at L-band from 1 to 2 GHz on 2017 September 3
and S-band from 2 to 4 GHz on 2017 September 6. For both observations, data were recorded in a new imaging mode in which the correlation (visibility) data are integrated separately into 20 pulse phase ranges (``bins'') based on the pulsar's initial timing ephemeris.  We de-dispersed and subtracted the mean, period-averaged signal in the visibility domain using {\tt sdmpy}\footnote{\url{http://github.com/demorest/sdmpy}}, then calibrated and imaged each bin individually using CASA\footnote{\url{http://casa.nrao.edu}}.  The mean-subtraction removes all continuous sources from the image, while the pulsar's signal remains as it is peaked at a small subset of phase bins. We then convolved the set of images versus bin with the pulsar's profile template, and recorded the maximum value of the convolution for each image pixel, effectively performing a matched filter.  The pulsar was not detectable at S-band, but the L-band matched filtered data resulted in a clear detection (see Figure~\ref{fig:VLA} right). 
From the matched-filter image, we find the position of PSR J1946$+$2052 to be 19:46:14.130(6) +20:52:24.64(9).
We fit the DD binary model~\citep{1985AIHS...43..107D,1986AIHS...44..263D} with the pulsar position fixed to that measured from the 
VLA localization using {\tt tempo}. The resulting timing solution is shown in Table~\ref{tab:timsol}.

\subsection{Multiwavelength analysis}
\label{sec:multiwave}
Once the localization was achieved, we examined multiple data
archives at the measured interferometric position and found a nearby 
source in IR (UKIDSS) and optical (IPHAS and SDSS) images. The nearby source is named J194614.14+205224.7
in the UKIDSS Galactic Plane Survey~\citep[UGPS][]{2008MNRAS.391..136L} and is $0\farcs175$ away from the VLA position of
J1946+2052. The IPHAS~\citep{2005MNRAS.362..753D} source is identified as J194614.14+205224.5 and is $0\farcs16$ away from the
position of J1946+2052. We calculated the probability of such a chance coincidence using the IPHAS 
source counts to be $\sim$0.002. However, in both catalogs the source is identified as a galaxy based
on the source being extended. We also checked and did not see any indication of a source at the location
of PSR J1946+2052 in 2MASS, DSS, GALEX, ROSAT, and Fermi archival data.

\begin{table*}
  \renewcommand{\arraystretch}{0.8}
  \begin{center} {\footnotesize
  \caption{Fitted and derived parameters for PSR J1946$+$2052.}
  \begin{tabular}{lc}
  \hline\hline
  \multicolumn{2}{c}{Measured Parameters} \\
  \hline
  Right ascension, $\alpha$ (J2000.0) \dotfill & {19:46:14.130(6)\tablenotemark{1}} \\
  Declination, $\delta$ (J2000.0) \dotfill & {20:52:24.64(9)\tablenotemark{1}} \\
  Pulse frequency, $\nu$ ($\mathrm{s^{-1}}$) \dotfill & 58.9616546384(5) \\
  First derivative of pulse frequency, $\dot \nu$ ($\mathrm{s^{-2}}$) \dotfill & $-$3.2(6)$\times 10^{-15}$\\
  Epoch (MJD) \dotfill & {57989.0}  \\
  Dispersion measure, DM ($\mathrm{pc\;cm^{-3}}$) \dotfill & 93.965(3) \\
  Ephemeris \dotfill & {DE436} \\
  Clock \dotfill & {TT(BIPM)} \\
  Span of Timing Data (MJD) \dotfill & 57953--58024                  \\
  RMS Residual ($\mu s$) \dotfill & 95.04 \\
  Binary model \dotfill & DD \\
  Orbital period, $P_\mathrm{b}$ (days) \dotfill & 0.07848804(1) \\
  Projected semimajor axis, $x$ (lt s) \dotfill & 1.154319(5) \\
  Orbital eccentricity, $e$ \dotfill & 0.063848(9) \\
  Epoch of periastron, $T_\mathrm{0}$ (MJD) \dotfill & 57989.002943(3) \\
  Longitude of periastron, $\omega$ (degrees) \dotfill & 132.88(1)\\
  Rate of periastron advance, $\dot \omega$ (degrees/yr) \dotfill & 25.6(3) \\
  1400 MHz mean flux density Arecibo ($\mathrm{\mu}$Jy) \dotfill & 62(6) \\
  1400 MHz mean flux density VLA ($\mathrm{\mu}$Jy) \dotfill & 84(15) \\
  \hline\hline
  \multicolumn{2}{c}{Derived Parameters} \\
  \hline
  Galactic latitude, $l$ (degrees) \dotfill & 57.66 \\
  Galactic longitude, $b$ (degrees) \dotfill & $-$1.98 \\
  DM-derived distance (NE2001), $d_\mathrm{DM}$ (kpc) \dotfill & 4.2 \\
  DM-derived distance (YMW16), $d_\mathrm{DM}$ (kpc) \dotfill & 3.5 \\
  Spin period, $P$ (s) \dotfill & 0.0169601753230(2)\\
  Spin period derivative, $\dot P$ \dotfill & 9(2)$\times$10$^{-19}$ \\
  Characteristic age, $\tau_\mathrm{c} = P/2 \dot{P}$ (Myr) \dotfill & 290 \\
  Surface magnetic field, $B_\mathrm{S} = 3 \times 10^{19}\sqrt{ P \dot P}$ ($10^{9}$G) \dotfill & 4 \\
  Spindown luminosity ($10^{32}$ erg/s)\dotfill  & 75 \\
  Mass function, $f_\mathrm{mass}$ ($M_\odot$) \dotfill & 0.268184(12) \\
  Total mass, $M_\mathrm{Total}$ ($M_\odot$) \dotfill & 2.50(4) \\
  \hline
\end{tabular} }
\end{center}
\tablecomments{Numbers in parentheses represent 1-$\sigma$ uncertainties from \texttt{tempo}, scaled for reduced $\chi^2 = 1$.}
\tablenotetext{1}{VLA positions, fixed in the {\tt tempo} fit.}
\label{tab:timsol}
\end{table*}

\section{Discussion}\label{sec:disc}

\subsection{Formation and nature of the system} \label{sec:formation}

Through the precise localization of PSR J1946$+$2052, the period derivative could be determined to be 9(2)$\times$10$^{-19} \, \rm s \, s^{-1}$
after  only 71 days of timing. Ignoring the effects from the Shklovskii effect and Galactic acceleration which are expected to be
very small for PSR J1946$+$2052, this suggests a 
characteristic age of 290\,Myr and surface dipole magnetic field of $4 \, \times \, 10^{9}\rm\,G$. The small
spin-down rate indicates the pulsar was recycled. The implied matter transfer from the companion progenitor would also have circularized the orbit. If the companion had evolved to a white dwarf the system would have retained a nearly
circular orbit, as observed for nearly all pulsar - white dwarf systems. 
We measure, however, a significant system eccentricity $e \, = \, 0.064$,
which requires a kick and/or sudden mass loss associated with the supernova (SN) of the progenitor of the companion. 
This evidence, coupled with the mass fraction, indicates that the companion is a NS, and the system a DNS.

In almost all aspects of its measured orbital parameters, the PSR J1946+2052 DNS resembles a further evolved version of the double pulsar J0737$-$3039A/B \citep{2003Natur.426..531B}. Below we discuss the implications of this similarity, assuming the companion mass is the same
as in the case of the double pulsar ($m_2 \, \sim \, 1.25 \, M_{\odot}$). In that
case, the constraint on the total mass of the system
(see Section~\ref{sec:total_mass}) yields $m_1 \, = \, 1.25 \, M_{\odot}$. 

Integrating the equations for the orbital decay back 290\,Myr
we 
derive firm upper limits on the orbital eccentricity and period at birth of
$e < 0.14$ and $P_b < 0.17$\,d.
As in the double pulsar \citep{2007MNRAS.379.1217L}, these indicate that
these systems had a small separation {\em before} the second SN.

Low eccentricities at birth imply small kicks associated with the
second SN. In the case of the double pulsar, the eccentricity,
proper motion, and misalignment between the spin and orbital axes 
($<3^\circ$, see \citealt{2013ApJ...767...85F}) are all very small.
That  conclusively limits the second SN kick to $\sim \, 70\, \rm km \, s^{-1}$
\citep[e.g.][]{2005PhRvL..94e1102P}. This small kick  suggests a close
binary interaction prior to the second SN, which stripped off the envelope of the
evolving secondary. Such ultra-stripped SNe
have smaller kicks and seem to produce lighter NSs \citep{2017ApJ...846..170T}.

If the companion of PSR~J1946+2052 has also originated in such a low-kick SN, then we should
expect the system to be similar to the double pulsar: a small value for $m_2$, a small peculiar velocity (the system would have a low velocity relative to the
local standard of rest, LSR) and a relatively close alignment between the spin axis and the orbital
angular momentum for PSR J1946$+$2052. 
As in PSR~J0737$-$3039A \citep{2013ApJ...767...85F}, we should not expect any pulse-profile changes due to geodetic precession. We are testing these predictions with continued observations.

\subsection{Measurement and prediction of Post-Keplerian parameters and system masses}
\label{sec:total_mass}
\begin{figure}
\includegraphics[trim=2.5cm 0cm 3cm 1cm, clip=true, width=1\columnwidth]{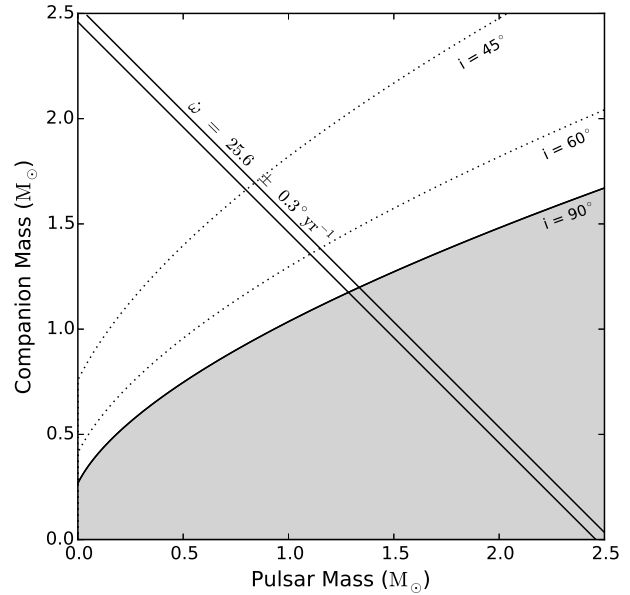}
\caption{The possible values for the mass of the pulsar and companion. The gray region
is not allowed due to the mass function. The black lines show the constraints due to the measurement of $\dot \omega$, as given by GR. The dotted lines represent varying orbital inclination angles ($i$).}
\label{fig:MpMc}
\end{figure}

Of all pulsars known in DNSs, PSR~J1946+2052 has the shortest orbital period. It also has the largest 
rate of advance of periastron, $\dot{\omega}\, = \, 25.6^\circ \rm yr^{-1}$.
Still, if this advance solely is due to  GR 
\citep{rob38,tw82}, the inferred total mass of $M_{\rm Total} \, = \, 2.50 \, \pm \, 0.04 \, M_{\odot}$ is 
{\bfref potentially less than the} lightest DNS known, PSR~J1411+2551 \citep{2017ApJ...851L..29M}.
From $M_{\rm Total}$ and the mass function  ($f\, = \, 0.268184(12) \, M_{\odot}$)
we derive an upper limit for the mass of the pulsar ($m_1\, < \, 1.31 \, M_{\odot}$)
and lower limit for the companion mass ($m_2\, > \, 1.18 \, M_{\odot}$); see Figure~\ref{fig:MpMc}.

{\bfref Without further PK parameters we cannot yet determine the individual NS masses.
For $m_1 \, = \, 1.25 \, M_{\odot}$ and $m_2 \, = \, 1.25 \, M_{\odot}$ we expect
an Einstein delay $\gamma \, = \,0.262\, \rm ms$. That is small compared to other DNS 
systems, but nevertheless simulations indicate that continued timing will measure
$\gamma$ with $\sim \, 10\%$ and $\sim 1\%$ relative uncertainty
by mid-2019 and mid-2025, respectively. This will allow a precise measurement
of both masses. Furthermore, for the masses assumed above, GR predicts an orbital decay
due to the emission of GWs of
$\dot{P}_b \, = \, -1.78\, \times \, 10^{-12}\,  \rm s \, s^{-1}$. This will be measured
with a relative uncertainty of $\sim 7.5 \%$ and $\sim 0.2 \%$ by mid-2019 and
mid-2025, respectively}.

\subsection{PSR~J1946+2052 as a gravitational laboratory}
{\bfref The measurement of $\dot{P}_b$} will be contaminated by two kinematic effects:
first, by the difference in Galactic acceleration of the Solar System and
the pulsar (\citealt{1995ApJ...441..429N}, estimated below using the Galactic parameters
from \citealt{2014ApJ...783..130R}) and second by the pulsar proper motion \citep{1970SvA....13..562S}.
These can be corrected once the distance is known.
{\bfref The NE2001~\citep{2002astro.ph..7156C} and YMW16~\citep{2017ApJ...835...29Y}
models predict distances of 4.2 and 3.5\,kpc, respectively}.
Given that and the faintness of the pulsar, 
a precise distance from VLBI or HI absorption (as done for PSR J1906+0746 in \citealt{2015ApJ...798..118V})
appears unlikely in the near future. Nevertheless, it is possible to estimate these kinematic contributions, 
assuming a pulsar in the LSR, with proper motion $\sim \, 6 \, \rm mas \, yr^{-1}$, over a range of distances of 4.2$\pm$1\,kpc. 
The sum of the kinematic contributions to $\dot{P}_b$
for $d \,= \,$3.2 and 5.2 kpc changes by only $+2.1/-2.6 \, \times \, 10^{-16} \, \rm s \, s^{-1}$.
With such a small uncertainty, the GR prediction for the orbital decay can be tested to
a precision of 0.015\%. This is one order of magnitude better than the 0.16\% test possible with the Hulse-Taylor pulsar \citep{2016ApJ...829...55W}. The quality of the 
PSR~J1946+2052 test will depend very significantly on its proper motion, making
its measurement an important
objective of future timing.


\subsection{Implications of PSR J1946$+$2052 on the DNS Merger Rate}
The large expected rate of orbital decay implies the system will merge quickly.
Indeed, based on the measured orbit and total mass, and the likely component masses,
the coalescence timescale is only 46 Myr. That is 
significantly shorter than the coalescence timescales of the double pulsar (85 Myr) and
PSR~J1757$-$1854 (76 Myr). The {\bfref current} GW luminosity of PSR~J1946$+$2052 is the largest of any known DNS:
$\sim$13\% of a solar luminosity (compared to 6.2\% for the double pulsar and 10.8\%
for PSR~J1757$-$1854).

Since PSR J1946+2052 strongly resembles a more evolved double pulsar system,
we assume they belong to similar populations of DNS binaries. We thus use
existing population models \citep[][]{2015MNRAS.448..928K} 
to calculate how many J0737$-$3039-like
binaries the PALFA survey can detect. We use PsrPopPy\footnote{\url{https://github.com/samb8s/PsrPopPy}} \citep[][]{2014MNRAS.439.2893B} to perform
the population synthesis and analysis.
        
The total number of these systems in the Galaxy is $N_{\rm pop} = 1500^{+4000}_{-1000}$
\citep[95\% confidence interval;][]{2015MNRAS.448..928K}. Using PsrPopPy, we generated a population 
of $N_{\rm pop}$ pulsars with the same spin period and orbital parameters as PSR J1946+2052. 
We found that given this population, the PALFA survey should have detected $2^{+5}_{-1}$ DNS systems
like J1946+2052, to date.
Therefore, the discovery of PSR J1946+2052 is predicted by the population models used for current 
merger rate estimates, and is unlikely to dramatically change the most recently published rate of $\mathcal{R}_\textrm{g} = 21^{+28}_{-14}$~Myr$^{-1}$ \cite[][]{2015MNRAS.448..928K}.

\subsection{Spin effects during the merger}

The small spin period, its derivative, and the relatively short coalescence time imply that the pulsar will
still be spinning rapidly when it merges. For braking indices between 0 and 3, the
values vary between 17.9 and 18.5\,ms.
For the larger spin period, the pulsar's dimensionless spin parameter
at merger will be given by
\begin{equation}
\chi \, = \, \frac{c}{G} \frac{2 \pi I }{m_1^2 P} = 0.032,
\end{equation}
where we have assumed $m_1 \, = \, 1.25 M_{\odot}$ and a moment of inertia
$I \, = 1.25\, \times \, 10^{45}\, \rm g \, cm^{2}$ 
\citep{2017arXiv171109226Z}. This is the
largest $\chi$ at merger for any pulsar in a confirmed DNS system including PSR~J1757$-$1854
\citep{2017arXiv171109226Z}. {\bfref This large spin parameter has implications} for the ability
to determine neutron star masses from the GW signals during NS-NS inspirals.
As shown in Table 1 and Figure 4 of \cite{2017PhRvL.119p1101A}, a constraint on $\chi$ is necessary
to precisely determine the masses of the individual NSs in a DNS merger.
With $| \chi | \, \leq \, 0.05$  the primary and secondary NS masses are $1.36$--$1.60\, M_{\odot}$
and $1.17$--$1.36\, M_{\odot}$, respectively; with $| \chi | \, \leq \, 0.89$ constraint
the limits are $1.36$--$2.26\, M_{\odot}$ and $0.86$--$1.36\, M_{\odot}$, respectively. 
Thus, knowing the plausible range of values for $\chi$ at merger is important for estimating masses 
from GW observations of DNS mergers.

\section*{Acknowledgements}
The Arecibo Observatory is operated by SRI International under a cooperative agreement 
with the National Science Foundation (NSF; AST-1100968), and in alliance with Ana G. M\'{e}ndez-Universidad
Metropolitana, and the Universities Space Research Association.
The National Radio Astronomy Observatory is a facility of the NSF operated under
cooperative agreement by Associated Universities, Inc.
K.S., A.B., S.C., J.M.C., P.D., M.A.M., S.M.R., and F.C. are (partially) supported by the NANOGrav Physics
Frontiers Center (NSF award 1430284). N.P. and M.A.M. are supported by NSF award number 1517003.
I.H.S. and V.M.K. both acknowledge NSERC Discovery Grants and the Canadian Institute for Advanced Research (CIFAR). V.M.K.
received further support from an NSERC Discovery Accelerator Supplement, NSERC's Harzberg Award, the
Canada Research Chairs Program, and the Lorne Trottier Chair in Astrophysics and Cosmology.
S.M.R. is a CIFAR Senior Fellow.
P.C.C.F., J.v.L., R.S.W. and J.W.T.H. gratefully acknowledge financial support by the European Research Council,
under the European Union's Seventh Framework Programme (FP/2007-2013) grant agreements
279702 (BEACON), 617199 (ALERT), 610058 (BLACKHOLECAM) and 337062 (DRAGNET), respectively.
P.C.C.F. further acknowledges support from the Max Planck Society; J.W.T.H. from a NWO Vidi
fellowship; J.S.D was supported by the NASA Fermi program; and
W.W.Z. by  the 
Chinese Academy of Science Pioneer Hundred Talents Program and the National Key R\&D Program of China No. 2017YFA0402600.
We thank Lijing Shao and Nobert Wex for stimulating discussions and useful suggestions.

\facility{Arecibo, EVLA, GBT}
\software{PRESTO, PSRCHIVE, TEMPO, PsrPopPy, sdmpy, CASA}

\bibliography{palfa_J1946}
\bibliographystyle{yahapj}

\end{document}